# Van der Waals forces control ferroelectric-antiferroelectric ordering in ABP$_2$X$_6$ laminar materials

Jeffrey R. Reimers,[*a] Sherif Abdulkader Tawfik [b] and Michael J. Ford [*b]

We show how van der Waals (vdW) forces outcompete covalent and ionic forces to control ferroelectric ordering in CuInP$_2$S$_6$ nanoflakes as well as in CuInP$_2$S$_6$ and CuBiP$_2$Se$_6$ crystals. While the self-assembly of these 2D layered materials is clearly controlled by vdW effects, this result indicates that the *internal* layer structure is also similarly controlled. Using up to 14 first-principles computational methods, we predict that the *bilayers* of both materials should be antiferroelectric. However, antiferroelectric nanoflakes and bulk materials are shown to embody two fundamentally different types of inter-layer interactions, with vdW forces strongly favouring one and strongly disfavouring the other compared to ferroelectric ordering. Strong specific vdW interactions involving the Cu atoms control this effect. *Thickness-dependent* significant cancellation of these two large opposing vdW contributions results in a small net effect that interacts with weak ionic contributions to control ferroelectric ordering.

## Introduction

Of recent recognition is the general principle that van der Waals forces can compete against covalent and ionic forces to control chemical bonding.[1] A critical application of this is the stabilization of gold surfaces and nanoparticles by Au(0)-thiyl bonding rather than the Au(I)-thiolate bonding believed for decades to dominate in these systems.[2] This effect is not restricted to nanotechnology but also leads to Fe(II)-thiyl versus Fe(II)-thiolate valence tautomerisation in cytochrome-P450 model compounds.[3] A classic catalytic system is the very strong binding of benzene to copper surfaces,[4] a system traditionally believed to involve Cu-C covalent bonding that instead now forms a paradigm for the understanding of strong van der Waals forces.[5] The van der Waals force is well known as being critical to the control of self assembly,[6, 7] hence strongly affecting many conceived devices made from 2D materials. Here, we show that strong specific van der Waals interactions associated with soft[8, 9] copper ions go beyond this to control *internal* ferroelectric ordering in polar laminar materials containing sheets of soft sulfur or selenium atoms. These forces prove to be sufficiently strong to, in certain cases, overturn the natural electrostatic forces driving ferroelectricity to make antiferroelectric structures – a somewhat rare effect that is only recently becoming understood.[10]

Bulk materials of the form ABP$_2$X$_6$ often take on a laminar shape and are of considerable current interest as possible ultrathin materials for use as ferroelectrics in memory storage and other devices. Bulk systems investigated in this context include: CuInP$_2$S$_6$,[11-17] CuBiP$_2$Se$_6$,[18] AgBiP$_2$Se$_6$,[18] AgBiP$_2$S$_6$,[18] CuCrP$_2$S$_6$,[19] and CuVP$_2$S$_6$.[20] However, materials of interest can take on any form with X as a chalcogenide S, Se, Te, etc., and A/B as either monovalent/trivalent or else divalent/divalent metal combination. Their laminar shape comes about as the materials contain [(PX$_3$)-(PX$_3$)]$^{4-}$ anions that assemble with their P-P bonds aligned like a bed of nails that is then covered by two parallel "sheets" of X atoms. Between the sheets, the A and B metal atoms embed into nominally octahedral holes formed between the X atoms of different anions. The property of interest is that the metal atoms may sit at the centre of these octahedral holes, making paraelectric structures with hexagonal symmetry that have no dipole moment, or else they can move to a side of the octahedral hole to become close to just one of the two X planes, creating local polarization orthogonal to the layer. Throughout a layer, these dipole moments can align ferroelectrically, antiferroelectrically, or randomly, giving rise to observed intra-layer phase transitions.[18]

If a single layer is internally ferroelectric, then adjacent layers can align either ferroelectrically or antiferroelectrically with respect to it, and it is this process with which we are primarily concerned herein. It is relevant to both bulk materials as well as to ultrathin materials, substances often called "nanoflakes". Nanoflakes can be made by exfoliating bulk materials[17] and possibly also by self-assembly of individually prepared layers. In both cases, layer stacking will be influenced by both inter-layer electrostatic forces as well as by van der Waals forces, making nanoflakes susceptible to external manipulation. What are the dominant features that control this stacking? If a layer can be thought of in the same way as a laboratory bar magnet, then the electrostatic force will favour ferroelectric alignment. Does the van der Waals force also show preference?

For ABP$_2$X$_6$ materials, recent experiments reveal results with quite different characteristics: CuInP$_2$S$_6$ bulk[11-13] and nanoflakes[17] are ferroelectric whereas the closely related material CuBiP$_2$Se$_6$ forms an antiferroelectric bulk polymorph.[18] In this work, we use first-principles calculations

---

[a.] *International Centre for Quantum and Molecular Structures and School of Physics, Shanghai University, Shanghai 200444, China.*
[b.] *School of Mathematical and Physical Sciences, University of Technology Sydney, Ultimo, New South Wales 2007, Australia.*
† Email Jeffrey.Reimers@uts.edu.au, reimers@shu.edu.cn, sherif.abbas@uts.edu.au, Mike.Ford@uts.edu.au

utilizing density-functional theory (DFT) to show how these features arise.

## Methods

All calculations are performed using VASP 5.4.1,[21] where the valence electrons are separated from the core by use of projector-augmented wave pseudopotentials (PAW).[22] The energy cut-off for the plane-wave basis functions was set at 500 eV for bulk solids but otherwise kept at the default values obtained using "PREC=HIGH". The energy tolerance for the electronic structure determinations was set at $10^{-7}$ eV to ensure accuracy. We use a k-space grids for the 4×4×1 for bilayers, 4×4×2 for two-layer solids, and 4×4×1 for six-layer solids, with expansion to double these sizes being shown to affect calculated energy differences by less than 1 meV. VASP performs periodic imaging in all three dimensions and so to model bilayers we introduced a vacuum region of ca. 15 Å between the periodic images in the direction normal to the bilayer. To minimize interactions between these periodic images we applied dipolar correction as implemented in VASP (by setting the "IDIPOL" to "TRUE") in early calculations. However, the effect was found to be negligible (< 1 meV) and so this sometimes troublesome procedure was no longer applied. As a second check on the calculation of ferroelectric interaction energies, calculations were performed on systems expanded three times in the $z$ direction, with average energies also varying by less than 1 meV.

Geometry optimizations were performed for all systems, terminating when the forces on all atoms fell below 0.01 eV/Å. For nanoflakes, in-plane hexagonal unit-cells of dimension 6.5532 Å for $CuBiP_2Se_6$[18] and 6.0955 Å for $CuInP_2Se_6$,[12] are used, whereas all lattice vectors are fully optimized for bulk solids. The starting structures for geometry optimizations were based on the properties of adjacent layers observed in these materials, with test calculations performed for alternate structures using the PBE-D3 method not realizing lower-energy alternatives. Hence, whenever possible, hexagonal symmetry and/or inversion symmetry was enforced in the calculations, with molecular alignments chosen so that the symmetry of the structure paralleled symmetry elements in the plane-wave basis set. To further increase numerical stability, the "ISYM=2" flag was used to enforce inversion symmetry of the electronic density matrix.

Many computational methods considered involve the raw PBE density functional plus this combined with various vdW corrections: Grimme's D2 empirical correction[23] (PBE-D2), Grimme's D3 empirical correction[24] in its original form without Becke-Johnson damping with PBE (PBE-D3) and revPBE[25] (revPBE-D3), the exchange-hole based correction of Steinmann and Corminboeuf[26] (PBE-dDsC), the Tkatchenko-Scheffler method[27] (PBE-TS), that with self-consistent screening (SCS)[28] (PBE-SCSTS), and that extended to make Tkatchenko's many-body dispersion method[29, 30] (MBD@rsSCS) (VASP flag "IVDW= 202"). The other computational methods are all based around the vdW density-functional approach of Dion et al..[31] This vdW correction is applied with the revPBE density functional[32] (revPBEvdW), the optPBE density functional[32] (optPBEvdW) and the optB88 density functional[32] (optB88-vdW). Also, in a form modified by Lee *et al.*, it is combined[33] with the BP86 density functional[34] (vdW-DF2). Finally, we also consider the raw HSE06 hybrid density functional[35] as well as that combined with D3.[36]

All optimized coordinates are reported in Electronic Supporting Information (ESI).

## Results and Discussion

### A. Ferroelectric order in $CuInP_2Se_6$ crystals and nanoflakes

Crystal structures of the bulk phase of $CuInP_2Se_6$ depict internally ferroelectrically ordered layers that ferroelectrically self-assemble with each other, producing ferroelectric crystals that can be exfoliated to produce ferroelectric nanoflakes.[11-13, 17] Its observed crystal structure is depicted in Fig. 1. Nominally hexagonal layers stack above each other in such a way that overall hexagonal symmetry is lost, with six symmetrically equivalent layers stacking vertically to form a unit that is periodic in the direction orthogonal to the layer plane. These six layers are named "a" to "f" in the figure. While the In atoms remain close to the centres of their octahedral holes, the Cu atoms move to take on nearly trigonal-planar arrangements within a single S sheet.

No antiferroelectric arrangement of the $CuInP_2Se_6$ has been observed, so we consider a possible form made simply by moving the Cu atoms from one S sheet in a layer to the other, allowing the In and other atoms to relax accordingly. In Fig. 1, we apply this transformation to alternate layers, converting the original ferroelectric crystal with a structure depicted as [abcdef] to the antiferroelectric structure [a′bc′de′f]. The same procedure could also have been applied to make an alternate antiferroelectric structure [ab′cd′ef′], but this is symmetrically equivalent and hence not shown. All optimized antiferroelectric structures embody inversion symmetry linking adjacent layers, so these structures have no net dipolar polarization. By sliding adjacent layers with respect to each other, many possible variants of the shown antiferroelectric structure are conceivable. The energy differences between such structures are subtle and will be considered elsewhere. Here we focus on a key feature of all antiferroelectric structures: as Fig. 1 shows, two very different types of inter-layer interactions are apparent depending on whether or not the Cu atoms face each other across the interface.

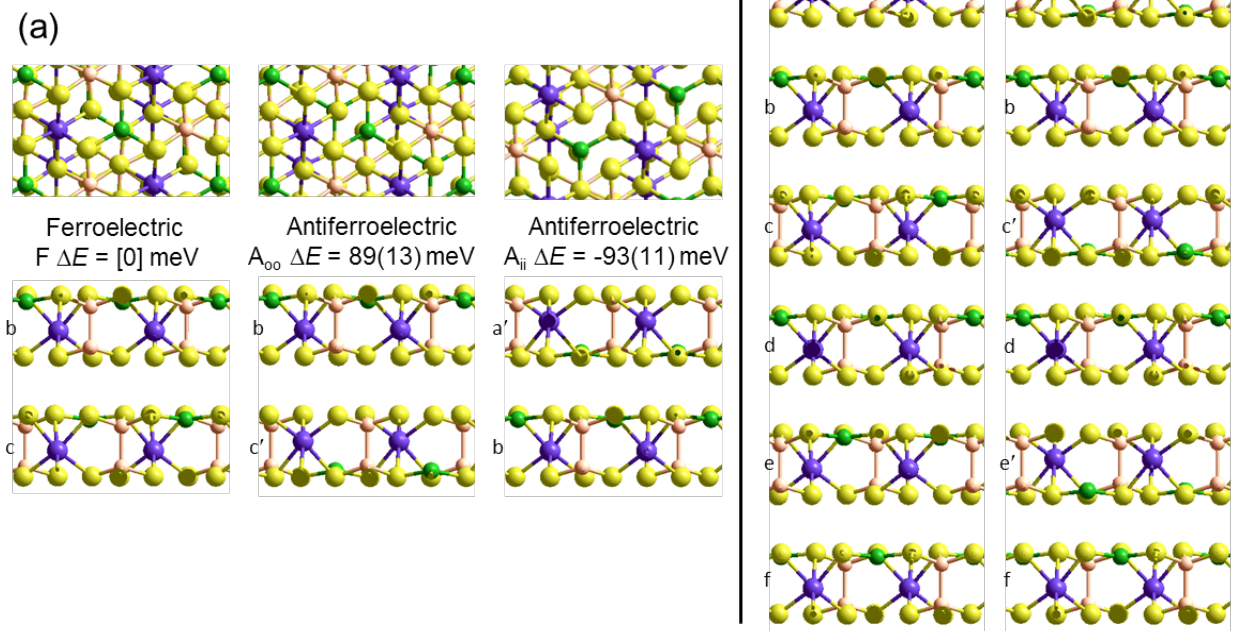

**Fig. 1** PBE-D3 optimized structures and relative energies for CuInP$_2$S$_6$ based on the observed ferroelectric structure of the bulk material: (b)- this bulk material with layers named [abcdef] and an antiferroelectric structure made from it by pushing the Cu and In atoms to the opposite side of their octahedral holes in alternate layers to make structure [a'bc'de'f], and (a) bilayers extracted from these solids. All six layers in the bulk ferroelectric structure are symmetrically equivalent to each other; similarly, the bilayer structures made by selecting any two adjacent layers in the bulk structure are all symmetrically equivalent, as are any (e.g., [b'c']) made from them by moving both Cu and In atoms to the opposite sides. Relative energies per bilayer are indicated as calculated by PBE-D3, with, in parenthesis, analogous energies calculated by PBE. Top and sectioned side views are shown for each structure: P- orange, Se- yellow, Cu- green, Bi- purple.

To highlight this effect, we consider possible bilayers derived from the (observed) ferroelectric and (envisaged) antiferroelectric bulk structures of CuInP$_2$S$_6$. The ferroelectric bilayer is named "F" and comprised of the layers [bc] extracted from the crystal. Extracting any other pair of layers, or interchanging the layer order to make [cb], produces symmetrically equivalent bilayers, so only one unique bilayer can be produced from the crystal in this manner. However, two unique antiferroelectric structures can be conceived, with one, named "A$_{ii}$", made by extracting layers [a'c], and the other, named "A$_{oo}$", made by extracting layers [bc'] instead. The A$_{ii}$ bilayers are so named as they have both Cu atoms adjacent to each other on the "inside" of the bilayer, whereas the A$_{oo}$ bilayers have them on the "outside", as far away from each other as is possible, see Fig. 1.

Figure 1 indicates the energies $\Delta E$ of the different antiferroelectric bulk and bilayer structures relative to that of the ferroelectric structures, obtained using two representative computational methods, (i) the PBE generalized-gradient approximation density functional[37], and (ii) this combined with the D3 dispersion correction in its original form (PBE-D3)[24]. The critical properties of these computational methods of relevance is that they similarly include covalent and ionic effects but only PBE-D3 embodies a realistic description of the van der Waals attraction term.

Both the PBE and PBE-D3 computational approaches predict that the observed ferroelectric bulk polymorph is more stable than its envisaged antiferroelectric variant by ca. 20 meV, a small but significant quantity. That the two approaches predict very similar outcomes is naively suggestive that electrostatic forces rather than van der Waals forces are most important in determining the electronic structure of laminal materials.

Alternatively, results for the bilayers presents a completely different picture of the nature of the intrinsic processes involved in self-assembly. PBE predicts that the antiferroelectric A$_{ii}$ and A$_{oo}$ bilayers are each about 6 meV less stable that the ferroelectric bilayer F, with all structures barely stable to spontaneous layer separation. As it predicts the bulk ferroelectric phase to be more stable that alternating $A_{oo}$ and $A_{ii}$ bilayer-type arrangements by 18 meV, a long-range favourable ferroelectric electrostatic interaction of 6 meV can be deduced. However, PBE-D3 predicts that $A_{ii}$ is more stable than F by 92 meV whereas $A_{oo}$ is less stable by 88 meV. Hence the van der Walls force is perceived as being very sensitive to the details of inter-layer arrangements, with the overall bulk structure emerging in part as a result of the way these contributions act to cancel each other out.

**Table 1.** Calculated bilayer structural energies differences $\Delta E_{Aii} = E_{Aii} - E_F$, between the lowest-energy antiferroelectric arrangement $A_{oo}$ and the ferroelectric structure F, in meV.

| method | CuInP$_2$S$_6$ | CuBiP$_2$Se$_6$ |
|---|---|---|
| PBE | 5 | -1 |
| PBE-D2 | -42 | -42 |
| PBE-D3 | -92 | -86 |
| revPBE-D3 | -81 | -63 |
| PBE-DFTdDsC | -33 | -36 |
| PBE-MBDrsSCS | -43 | -45 |
| PBE-SCSTS | -100 | -76 |
| PBE-TS | -83 | -87 |
| revPBEvdW | -22 | -26 |
| optB88vdW | -40 | -40 |
| optPBEvdW | -30 | -31 |
| vdWDF2 | -26 | -23 |
| HSE06 | -15 | -5 |
| HSE06-D3 | -59 | -64 |

Next we consider how robust this bilayer result is likely to be considering possible variations and improvements in the computational methods used. Table 1 shows results for the $A_{ii}$ bilayer energy with respect to the F bilayer energy obtained using the above two computational methods plus 12 others. In one case, the PBE generalize-gradient functional is replaced with the HSE06[35] hybrid functional, providing a much enhanced description of electron exchange. The small PBE preference of 5 meV for the ferroelectric structure compared to $A_{ii}$ is reversed by a 15 meV preference for $A_{ii}$, but adding D3 provides an extra 44 meV preference. This change is only half that (97 meV) predicted when D3 is added to PBE, but the general scenario remains unchanged. The 10 other methods utilized in Table 1 all involve variations of the PBE density functional combined with different dispersion approaches. In terms of magnitude of the effect, these approaches show quite a range with the total preference for $A_{ii}$ ranging from 27 to 105 meV – a significant range but again the qualitative effect remains preserved. Understanding these method differences is a topic for ongoing research, but a noteworthy results is the net stabilization of 81 meV predicted by revPBE-D3,[25] a method recently highlighted as one of the most robust methods of its type across all types of chemical scenarios.[38] Another is the value of 43 meV predicted by the many-body dispersion[29, 30] (MBD) method, an advanced treatment of dispersion that, to treat conductors better, does not assume

pairwise additivity;[39] it is also known to be widely robust to chemical scenario.[5]

Lastly, we consider the robustness of the main result considering subtle effects not normally considered in energetics analyses of materials. Calculations of the zero-point energy difference between the two structures at the PBE-D3 level indicated a change of less than 1 meV. Also, calculations of the room-temperature Helmholtz free energy change[40] at the PBE-D3 level indicate corrections of just -3 meV. Therefore the best-possible description from first principles calculations available at the moment is that bilayers of CuInP$_2$S$_6$ should be antiferroelectric.

To demonstrate that this prediction is consistent with the observation of ferroelectric nanoflakes of CuInP$_2$S$_6$, we consider what happens when a third layer is added to an existing $A_{ii}$ bilayer, say by spontaneous self-assembly. As shown in Fig. 2, a third layer can be added in one of two ways, making structure $A_{ii}$F, in which the new layer adds ferroelectrically to its neighbour, as well as structure $A_{ii}A_{oo}$, in which it adds antiferroelectrically. PBE-D3 calculations indicate that $A_{ii}$F is more stable than $A_{ii}A_{oo}$ by 88 meV, paralleling the previous result found for bilayer formation. Making a tetralayer system from this trilayer then proceeds analogously, with PBE-D3 predicting that $A_{ii}$FF is 87 meV more stable than $A_{ii}$FA$_{oo}$. Taken together, these results suggest that the energies of antiferroelectric multi-layer systems can be usefully described simply in terms of sums of bilayer interactions. The key prediction here is that nanoflakes grown under such conditions favouring sequential layer deposition increasingly develop more and more ferroelectric character as the number of layers increases.

However, if instead two $A_{ii}$ bilayers self-assemble to make a tetralayer, then the PBE-D3 calculations predict the resultant structure, $A_{ii}A_{oo}A_{ii}$, to be 8 meV more stable than $A_{ii}$FF, making the antiferroelectric structure the thermodynamically most stable one. This result is also easily understood in terms of pairwise inter-layer interactions. In Figs. 1 and 2, all results for bilayers, trilayers, and tetralayers were obtained constraining the in-plane lattice vectors to reference values. This minimizes the variables changing between calculations, allowing focus to be placed on key features such as the change in the van der Waals interactions between structures. Alternatively, the shown bulk structures are at fully relaxed lattices obtained individually for each computational method used, allowing focus to be placed on the small energy differences that in the end determine which structure is observed and which is not. Calculations of the bulk materials at the reference lattice vectors indicate energy changes of only 5 meV, however, so lattice relaxation does not have a profound effect. Also, the energies of the ferroelectric crystals at these reference geometries are 26 meV more stable compared to the antiferroelectric structures than one would expect simply by adding bilayer energies, an effect arising from long-range attractive dipole-dipole interactions present in ferroelectric structures. Proper treatment of these effects, plus use of the most accurate computational methods possible, will be critical to making robust predictions for properties of specific nanoflakes. For example, the calculated 8 meV energy difference between A$_{ii}$FF and A$_{ii}$A$_{oo}$A$_{ii}$ tetralayers is too small and too susceptible to method variations for this to be regarded as a robust prediction of tetralayer properties.

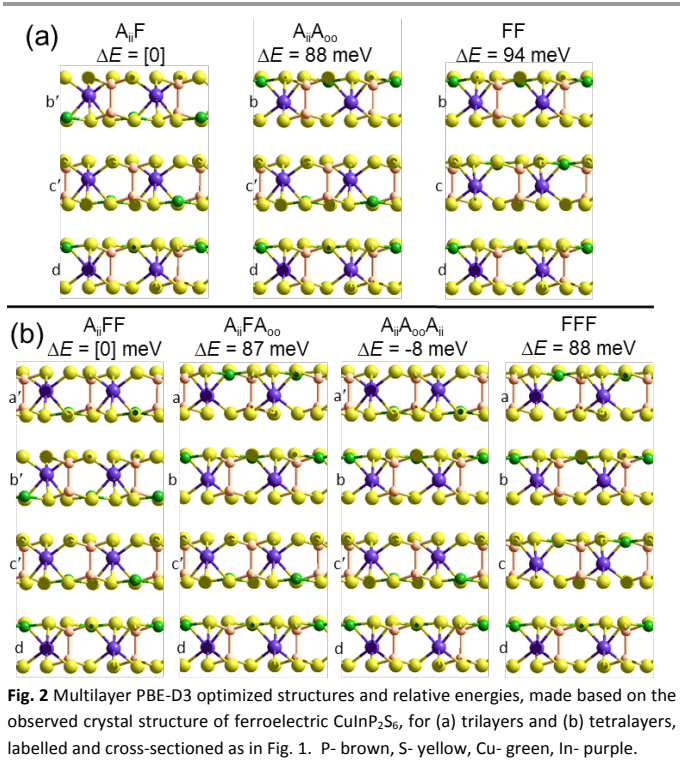

**Fig. 2** Multilayer PBE-D3 optimized structures and relative energies, made based on the observed crystal structure of ferroelectric CuInP$_2$S$_6$, for (a) trilayers and (b) tetralayers, labelled and cross-sectioned as in Fig. 1. P- brown, S- yellow, Cu- green, In- purple.

Nevertheless, robust predictions from the calculations are that CuInP$_2$S$_6$ should be antiferroelectric, that under kinetic control subsequent layers will add ferroelectrically to an initial antiferroelectric bilayer, and that ferroelectric CuInP$_2$S$_6$ structures are intrinsically more stable than antiferroelectric ones for large nanoflakes. Given this, then at some point during sequential layer growth, the point will come at which kinetically produced structures transform into their thermodynamically more stable counterparts. In general, the issue of kinetic verses thermodynamics control in molecular self-assembled monolayer production is of considerable currently of interest, with new experimental and computational methods recently developed[41] and reviewed.[1] In the experimentally observed nanoflakes,[17] layers were exfoliated from ferroelectric bulk material. Naively, one would expect such a high-energy exfoliation process to provide ample opportunity for the most thermodynamically stable products to be formed.

Through consideration of the details of the structures of the observed[17] nanoflakes, the calculated results provide a basis for understanding. These nanoflakes were ca. 1 μm in size with ca. half their area being 3 layers high, a quarter 4 layers high, and the remaining quarter 2 layers high. Trilayer structures, and indeed all structures containing an odd number of internally ferroelectrically ordered layers, must have net dipole polarization, the issue being whether the net polarization corresponds to that of just a single layer or else multiples of this up to the full ferroelectric order of the nanoflake. When only a small amount of a fourth layer is added to a trilayer system, the energetics indicate that it can only add ferroelectrically. The fact that antiferroelectric structures come in two fundamental types and that these must be combined in multi-layer systems explains the experimental observations despite predictions that bilayers should be antiferroelectric.

## B. Antiferroelectric order in CuBiP$_2$Se$_6$ crystal

This material is observed at low temperature to comprise an antiferroelectric combination of ferroelectric monolayers, see Fig. 3.[18] It differs from CuInP$_2$S$_6$ primarily through the horizontal alignment of the layers, with the alignment in CuBiP$_2$Se$_6$ crystals such that the hexagonal symmetry of each layer is maintained throughout the crystal. Again, six layers are present in the smallest unit cell that shows periodicity in the direction normal to the layers, but for CuBiP$_2$Se$_6$ these consists of layer pairs that are replicated each three times. The layers in the observed crystal are labelled [ab'cd'ef'] using the same notation as that already applied in Figs. 1 and 2. By simply transferring Cu atoms within octahedral holes, a ferroelectric alternative is envisaged as [abcdef] shown in Fig. 3; this is symmetrically equivalent to the other structure makeable in this fashion, [a'b'c'd'e'f']. From the observed antiferroelectric structure, two fundamentally different bilayer types can be extracted as A$_{ii}$ (e.g., [b'c] shown in Fig. 3) and as A$_{oo}$ (e.g., [ab] shown in Fig. 3). In contrast to bilayers built based on CuInP$_2$S$_6$, [bc'] is inequivalent to [ab'] while [a'b] is inequivalent to [bc']; hence these other possibilities are also shown in Fig. 3 and labelled A'$_{oo}$ and A'$_{ii}$, respectively. Also, two different types of ferroelectric bilayers can be simply constructed from the observed antiferroelectric crystal structure, with [ab] and [bc] shown as examples in the figure and named F and F', respectively. While PBE-D3 calculations indicate that these ferroelectric variants have very similar energies, large differences are again predicted between antiferroelectric forms, and we focus on this effect.

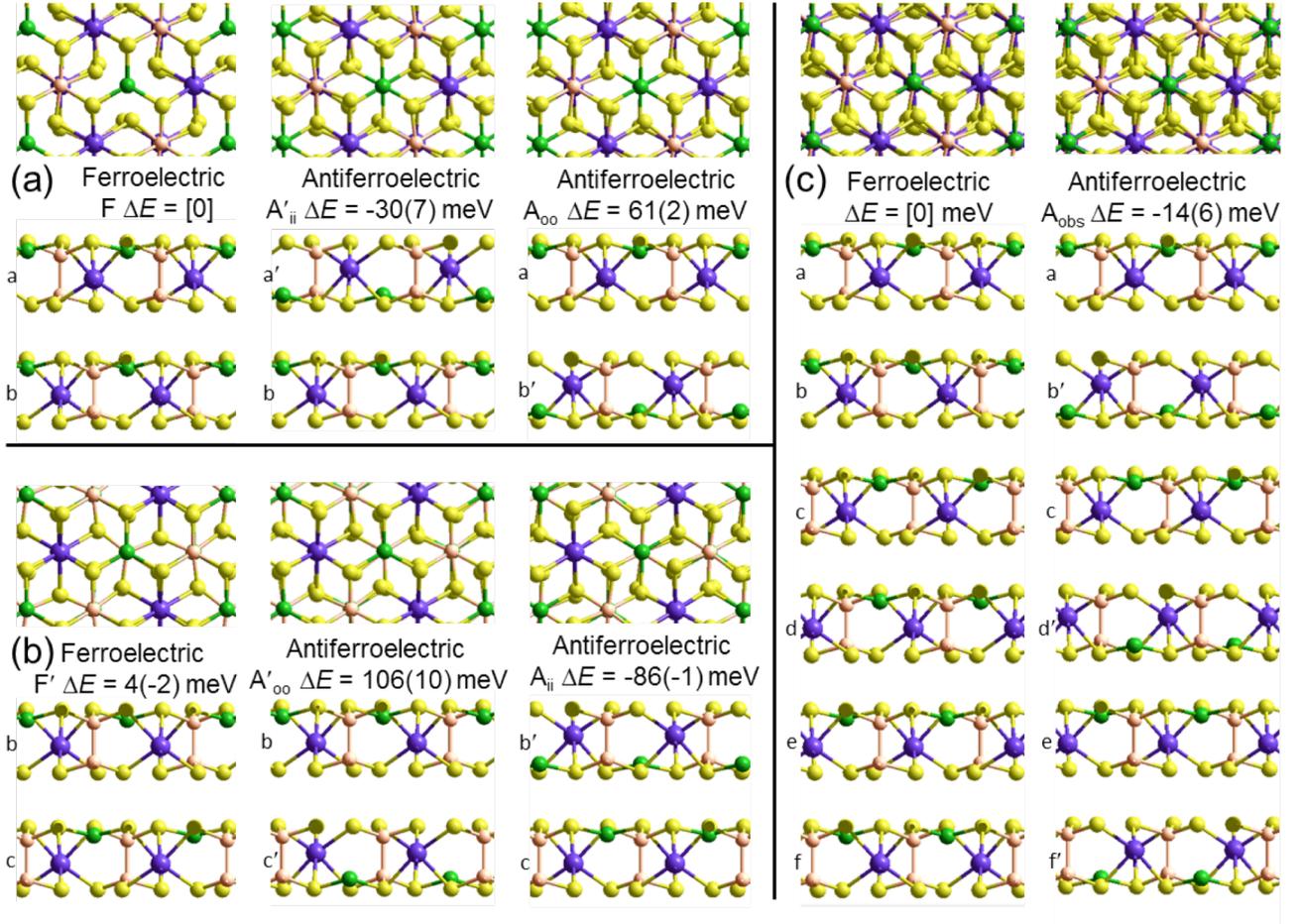

**Fig. 3** PBE-D3 optimized structures and relative energies for CuBiP₂Se₆ based on the observed antiferroelectric structure of the bulk material: (c)- this bulk material with layers named [ab'cd'ef'] and a ferroelectric structure made from it by pushing the Cu and Bi atoms to the opposite side of their octahedral holes in alternate layers to make structure [abcdef], (a) bilayers extracted from the top two layers of these solids, and (b) bilayers extracted from their second and third layers. The shown ferroelectric bilayer structures [ab] and [bc] are symmetrically equivalent to the unshown structures [a'b'] and [b'c'], respectively, while for the bulk solid the unshown structure [a'b'c'd'e'f'] is equivalent to [abcdef]; also, in the bulk solids, the later layer pairs are all symmetrically equivalent to the first pair. Relative energies per bilayer are indicated as calculated by PBE-D3, with, in parenthesis, analogous energies calculated by PBE. Top and sectioned side views are shown for each structure: P- orange, Se- yellow, Cu- green, Bi- purple.

The results shown in Fig. 3 for PBE and PBE-D3, supported by those obtained using the other methods shown in Table 1, indicate that variations in covalent and ionic bonding poorly discriminate between bilayer structures, whereas dispersion again has a profound effect. As for CuInP₂S₆, the attractive arrangement of dipoles in a ferroelectric bulk material stabilizes this form by ca. 7 meV, a small effect compared to the magnitude of the differential van der Waals forces. However, the significant difference to CuInP₂S₆ is that for it the stabilization and destabilization of adjacent bilayer interfaces present in the bulk material cancel, the (PBE-D3) stabilization of A_ii by 86 meV overpowers the destabilization of A_oo by 61 meV to provide the driving force for the formation of the observed antiferroelectric crystal.

**C. The van der Waals forces that control ferroelectric ordering**

Table 2 explores the origins of the differential van der Waals bonding between the F, A_ii, and A_oo structures, listing the van der Waals contributions coming from the D3 correction to CuBiP₂Se₆ and CuInP₂S₆ bilayer energies. The D3 correction takes on a "pairwise additive" form involving summation of London dispersion energies between every pair of atoms in the system.[5, 24, 39] Hence, in principle it is possible to decompose this contribution into terms arising from constituent atoms, and here we do this considering contributions from Cu to Cu interactions, Cu to non-Cu "$\overline{Cu}$" interactions, and non-Cu to non-Cu interactions using:

$$E^{D3} = \sum_{i \neq j}^{Cu} E_{ij}^{D3} + \sum_{i}^{Cu}\sum_{j}^{\overline{Cu}} E_{ij}^{D3} + \sum_{i \neq j}^{\overline{Cu}} E_{ij}^{D3}$$

To do this, we evaluate the D3 energy for subsystems containing either only Cu atoms or else no Cu atoms and compare the results to the full D3 energy of the system. This procedure is somewhat approximate as D3 is strictly not purely pairwise additive, with the van der Waals properties of each atom determined in the context of its environment, but the results will be qualitatively indicative of the role of copper in controlling inter-layer van der Waals interactions.

**Table 2** Analysis of the total D3-calculated dispersion energy $E^{D3}$, in eV, for F, A$_{ii}$, and A$_{oo}$ ABP$_2$X$_6$ bilayers, partitioned into contributions involving Cu atoms (Cu) and those not involving Cu atoms ($\overline{Cu}$); dispersion energy differences $\Delta E^{D3}$, in meV, from the F structures are also listed, as well as the percentage of the total dispersion energy arising from contributions involving Cu.

| | $E_F^{D3}$ | $E_{Aii}^{D3}$ | $E_{Aoo}^{D3}$ | $\Delta E_{Aii}^{D3}$ | $\Delta E_{Aoo}^{D3}$ |
|---|---|---|---|---|---|
| | CuBiP$_2$Se$_6$ | | | | |
| total | -4.04 | -4.13 | -3.96 | -83 | 84 |
| $\overline{Cu}$-$\overline{Cu}$ | -2.24 | -2.21 | -2.25 | 30 | -16 |
| Cu-$\overline{Cu}$ | -1.78 | -1.87 | -1.69 | -86 | 95 |
| Cu-Cu | -0.03 | -0.05 | -0.02 | -27 | 4 |
| % Cu | 45 | 47 | 43 | 136 | 119 |
| | CuInP$_2$S$_6$ | | | | |
| total | -3.66 | -3.75 | -3.56 | -95 | 99 |
| $\overline{Cu}$-$\overline{Cu}$ | -2.57 | -2.21 | -2.59 | 30 | -16 |
| Cu-$\overline{Cu}$ | -1.04 | -1.12 | -0.93 | -72 | 110 |
| Cu-Cu | -0.04 | -0.09 | -0.04 | -53 | 5 |
| % Cu | 30 | 32 | 27 | 131 | 116 |

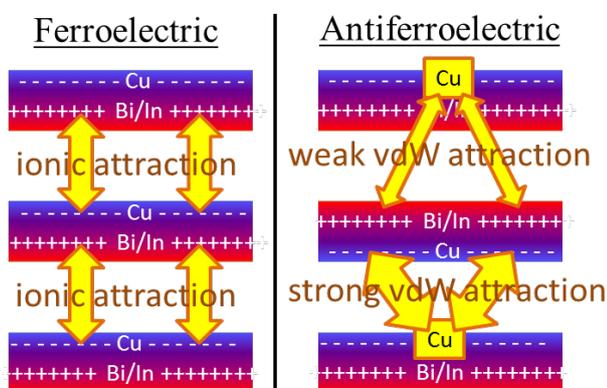

**Fig. 4** The primary ionic and van der Waals (vdW) interactions between layers of ABP$_2$X$_6$ laminar materials. The ionic forces are weak but provide net stabilization of ferroelectric polymorphs compared to antiferroelectric ones, whereas the differences in inter-layer van der Waals interactions focus on the interaction of the Cu atoms with the electron-rich parts of the neighbouring bilayer, strongly stabilizing A$_{ii}$ bilayers while destabilizing A$_{oo}$ bilayers compared to ferroelectric ones.

Table 2 shows that the total D3 contributions to bilayer energies range from near -3.6 eV for CuBiP$_2$Se$_6$ to near -4.0 eV for CuBiP$_2$Se$_6$, the difference arising owing to the higher polarizabilities of Se and Bi compared to S and In. For CuBiP$_2$Se$_6$, c.a. 45% of the dispersion energy arises through interactions with Cu atoms, reducing to ca. 30% for CuBiP$_2$Se$_6$. However, in terms of the D3 energy differences between the A$_{ii}$ or A$_{oo}$ structures and F, $\Delta E^{D3}$, in both cases the Cu-involving contribution is ca. 125% of the whole, with the $\overline{Cu}$ – $\overline{Cu}$ contribution opposing it but having only one fifth of the magnitude. The direct Cu–Cu van der Waals interaction contributes only one sixth to one third of this total, with the remainder coming from Cu–$\overline{Cu}$ interactions. However, both types of Cu contributions are much more attractive across A$_{ii}$ interfaces than they are across A$_{oo}$ ones.

Why the copper van der Waals forces are strongest over A$_{ii}$ junctions and weakest over A$_{oo}$ ones is sketched in Fig. 4. The most important van der Waals interactions involve the Cu of one layer interacting with the Cu and the electron-rich part of its neighbour. The magnitude of such interactions clearly scale in the order A$_{ii}$ > F > A$_{oo}$. Also in the figure the basic ionic inter-layer interactions are also sketched, clearly favouring F over A$_{ii}$ and A$_{oo}$. It is the balance achieved between the weak ionic forces and the strong but opposing van der Waals forces that controls the ferroelectric ordering in ABP$_2$X$_6$ laminar materials.

## Conclusions

What is revealed for both ABP$_2$X$_6$ laminar materials considered is that antiferroelectric structures intrinsically involve two different types of inter-layer interactions with significantly different van der Waals forces, one type being much more attractive than the van der Waals force acting between ferroelectric structures while the other is much less attractive. The difference between these two van der Waals bonding effects compared to the weak dipole-dipole stabilization energy operative in ferroelectric structures then controls the properties of crystals and nanoflakes.

Realizing the nature of the balance of forces controlling the structure of ABP$_2$X$_6$ laminar materials will facilitate new synthetic strategies for designing desirable materials, as well as new measurements to exploit bilayer properties. This occurs as external electric fields can manipulate the dipole-dipole interaction, with controlling field strengths being determined by the differential van der Waals interaction.

In general, the role of van der Waals forces and the manifold ways they can compete with traditional ionic and covalent bonding scenarios is a topic of great current interest,[1] with the results found here for ABP$_2$X$_6$ laminar materials providing another example. van der Waals forces not only effect self-assembly but also can control localized chemical structure.

## Conflicts of interest

There are no conflicts to declare.

## Acknowledgements

This work was supported by resources provided by the National Computational Infrastructure (NCI), and Pawsey Supercomputing Centre with funding from the Australian Government and the Government of Western Australia, as well as Chinese NSF Grant #1167040630. This work is also supported by the Australian Research Council grant DP16010130.

## Notes and references

**TOC graphic**

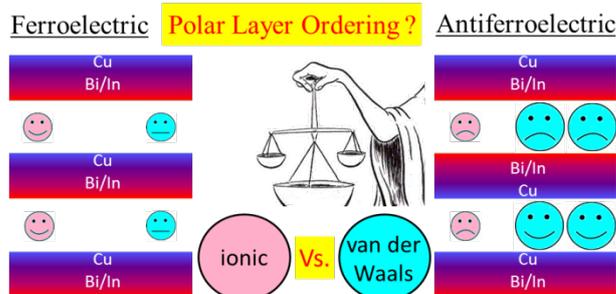